**A Deep Learning-Based Method for Automatic Segmentation of Proximal Femur from Quantitative Computed Tomography Images**

Chen Zhao[1#], Joyce H. Keyak[2#], Jinshan Tang[1], Tadashi S. Kaneko[3], Sundeep Khosla[4], Shreyasee Amin[5], Elizabeth J. Atkinson[6], Lan-Juan Zhao[7], Michael J. Serou[8], Chaoyang Zhang[9], Hui Shen[7], Hong-Wen Deng[7*], Weihua Zhou[1*]

1. College of Computing, Michigan Technological University, Houghton, MI, 49931, USA
2. Department of Radiological Sciences, Department of Mechanical and Aerospace Engineering, Department of Biomedical Engineering, and Chao Family Comprehensive Cancer Center, University of California, Irvine, Irvine, CA, 92868, USA
3. Department of Radiological Sciences, University of California, Irvine, Irvine, CA, 92868, USA
4. Division of Endocrinology, Department of Medicine, Mayo Clinic, Rochester, MN, USA
5. Division of Epidemiology, Department of Health Sciences Research, and Division of Rheumatology, Department of Medicine, Mayo Clinic, Rochester, MN, USA
6. Division of Biomedical Statistics and Informatics, Department of Health Sciences Research, Mayo Clinic, Rochester, MN, USA
7. Tulane Center of Bioinformatics and Genomics, Tulane University School of Public Health and Tropical Medicine, New Orleans, LA, 70112, USA
8. Department of Radiology, Tulane University School of Medicine, New Orleans, LA 70112, USA
9. School of Computing Sciences and Computer Engineering, University of Southern Mississippi, Hattiesburg, MS, 39406, USA

#, Co-first authors; *, Co-corresponding authors.

#Authors for Correspondence:

Weihua Zhou, PhD

College of Computing, Michigan Technological University,

1400 Townsend Dr, Houghton, MI, 49931, USA

E-Mail: whzhou@mtu.edu

Or

Hong-Wen Deng, Ph.D.

Tulane Center for Bioinformatics and Genomics

Department of Biostatistics and Data Science

Tulane School of Public Health and Tropical Medicine

1440 Canal Street, Suite 1610, New Orleans, LA 70112, USA

Email: hdeng2@tulane.edu

**Abstract**

**Purpose**: Proximal femur image analyses based on quantitative computed tomography (QCT) provide a method to quantify the bone density and evaluate osteoporosis and risk of fracture. We aim to develop a deep-learning-based method for automatic proximal femur segmentation.

**Methods and Materials**: We developed a 3D image segmentation method based on V-Net, an end-to-end fully convolutional neural network (CNN), to extract the proximal femur QCT images automatically. The proposed V-net methodology adopts a compound loss function, which includes a Dice loss and a L2 regularizer. We performed experiments to evaluate the effectiveness of the proposed segmentation method. In the experiments, a QCT dataset which included 397 QCT subjects was used. For the QCT image of each subject, the ground truth for the proximal femur was delineated by a well-trained scientist. During the experiments for the entire cohort then for male and female subjects separately, 90% of the subjects were used in 10-fold cross-validation for training and internal validation, and to select the optimal parameters of the proposed models; the rest of the subjects were used to evaluate the performance of models.

**Results:** Visual comparison demonstrated high agreement between the model prediction and ground truth contours of the proximal femur portion of the QCT images. In the entire cohort, the proposed model achieved a Dice score of 0.9815, a sensitivity of 0.9852 and a specificity of 0.9992. In addition, an $R^2$ score of 0.9956 (p<0.001) was obtained when comparing the volumes measured by our model prediction with the ground truth.

**Conclusion:** This method shows a great promise for clinical application to QCT and QCT-based finite element analysis of the proximal femur for evaluating osteoporosis and hip fracture risk.

## 1. Introduction

Osteoporosis is an initially silent bone disease characterized by fractures of the hip (the proximal femur), spine or wrist. Hip fractures are particularly debilitating and difficult to predict [1]. Quantitative computed tomography (QCT) is used to evaluate osteoporosis and risk of hip fracture by quantifying bone density and geometry of regions within the proximal femur [2, 6]. QCT is also combined with a structural analysis technique called finite element (FE) modeling to compute the force required to cause fracture of the proximal femur when the bone is subjected to specific loading conditions [3][4]. Although valuable research has been performed using this methodology, studies of large cohorts are hindered by the need for segmentation of the proximal femur, which typically involves a semi-automated technique requiring time consuming user monitoring and manual intervention. If a fully automated method of segmentation existed, large cohorts of subjects could be analyzed to facilitate new discoveries, such as those involving artificial intelligence.

Before the application of deep learning techniques, proximal femur segmentation was performed using statistical and semi-automatic methods in which the segmentation procedure requires user intervention to obtain precision, such as the primitive shape recognition methods [5], atlas-based segmentation [6] [7], graph-cut approaches [8], active model [9], and statistical shape models [10]. Even though user intervention guarantees precision of the segmentation results, it is tedious, and well-trained personnel are required. Several deep-learning based methods have been proposed. Zeng et. al proposed a 3D U-Net with multi-level deep supervision for automatic proximal femur segmentation on 3D MR images of femoroacetabular impingement [11]. Fang et. al used 3D feature-enhanced modules, including edge detection and multi-scale feature fusion modules to automatically extract the proximal femur from CT images [12]. However, existing deep-learning based methods have not achieved satisfactory results, which motivates us to develop a more precise algorithm to automatically segment the proximal femur.

In this paper, a deep-learning based method using a V-Net framework to automatically extract the proximal femur from QCT images is developed and validated.

## 2. Methods and Materials

### 2.1 Subject and Image Data

We obtained anonymized QCT scans of the hips of 216 women and 181 men [mean age (range): 60.9 (27 to 90) years, and 59.9 (28 to 90) years, respectively] from the year 6 follow-up visit of a previous longitudinal study of an age-stratified random sample of Rochester, MN residents [13]. The subjects included here are a subset of that study and were selected for analysis in a previous FE study [14]. Subjects previously provided written informed consent that extended to the analyses presented here. Ninety-five percent of the men and 99% of the women were white.

The QCT scans (Siemens, Sensation 64, 120kVp; 2-mm-thick slices; pixel size, 0.742 mm to 0.977 mm; convolution kernel, B30s; 512 x 512 matrix) were converted to 3-mm-thick slices for consistency and comparison with previous FE studies [15]. Fourier interpolation was performed over groups of three contiguous 2-mm-thick images to obtain six 1mm-thick contiguous images, followed by decimation (by averaging voxels in successive slices) to create two 3-mm-thick contiguous images.

To obtain the ground truth, the left proximal femur was segmented by one of the authors (TSK) using thresholding combined with an edge following algorithm [16] that was integrated with in-house software to create a user-interactive semi-automated segmentation procedure.

### 2.2 Data Acquisition and Data Pre-processing

During processing, the resolution of a single QCT slice was 512×512 with 3 mm slice thickness and pixel size of 0.8 mm. Only the ground truth of the left proximal femur, which is on the right side of each slice, was annotated, so we manually cropped a rectangular area in the middle-right part of the original image to train our model. The number of slices in the patients ranged from 37 to 95. The cropping process is shown in Figure 1.

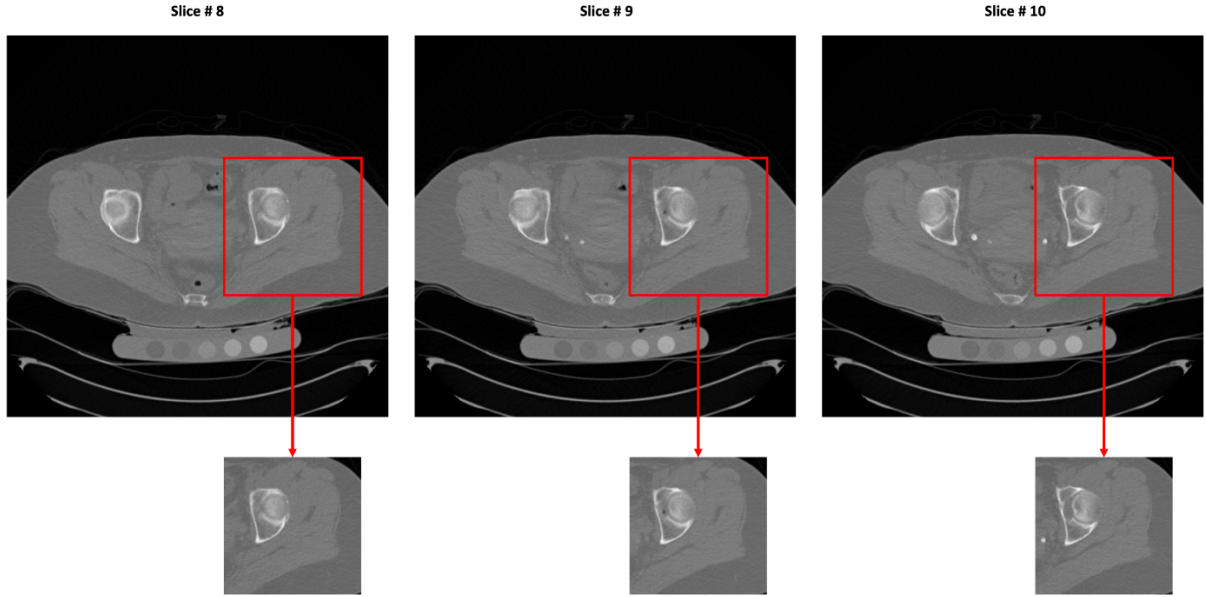

Figure 1. Cropping process for generating the input volume

Examples of cropped QCT slices with corresponding contours and the proximal femur region of interest (RoI) are shown in Figure 2.

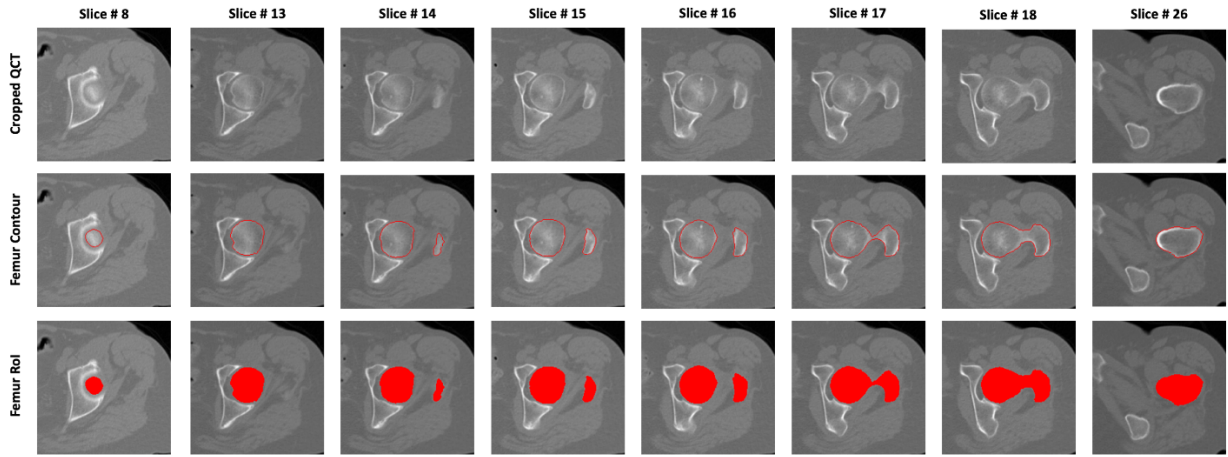

Figure 2. Examples of QCT slices (top row) with corresponding femur contours (middle row) and RoI (bottom row).

## 2.3 Proposed Deep Learning Model

The proposed femur segmentation method consisted of a volume-based deep learning network - a 3D end-to-end V-Net. The input of the model is a cropped QCT image volume with a size of 192×192×32 voxels. In the model, the 3D V-Net architecture is employed to encode the latent and high-level features of the inputted raw QCT images and decode the features to generate the final segmentation results. After feeding the cropped volume of a QCT image into the 3D V-Net, a Dice loss, which computes the discrepancy between the generated segmentation result and manually delineated ground truth (gold standard), is used for evaluating the current performance of the model. After that, the gradient of the loss is computed and back-propagation through the whole flow of the V-Net is employed to fine-tune the model parameters during the training epochs. The workflow of the proposed 3D V-Net is depicted in Figure 3.

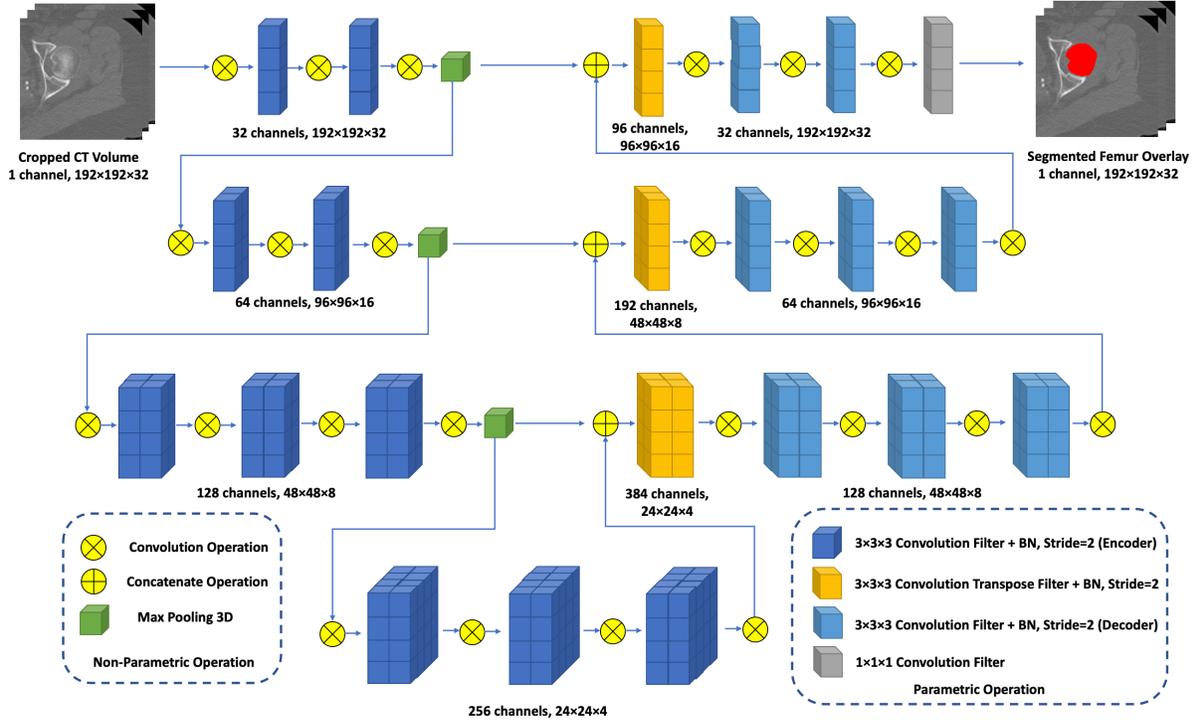

Figure 3. Schematic diagram of the proposed V-Net for femur segmentation.

As shown in Figure 3, the designed V-Net architecture of femur segmentation contains an encoder and a decoder. In detail, the encoder of the V-Net contains three down-sampling layers implemented using max pooling operations, and the symmetric decoder consists of three up-sampling layers implemented using transpose convolution operations. Each feature map after the pooling operation is fed into the decoder through the skip-connection operation, and it is concatenated with the corresponding feature map in the decoder. The number of feature maps in each convolution block is demonstrated in Figure 3. Before the model output layer is created, a convolutional layer with 1×1×1 kernel is employed to transfer the feature maps into a probability image with a resolution of 192×192×32. In addition, before the last convolutional layer is created, we add a dropout layer to prevent the model from overfitting [17]. The output of each voxel of the proposed V-Net is the probability that the particular voxel is contained within the proximal femur. Finally, the OTSU algorithm [18] is employed to convert this probability map into binary segmentation results where '1' represents a voxel in the femur, and '0' indicates the background.

### 2.4. Loss Function and Optimization Strategy

We employed a compound loss function that consists of a Dice loss and a L2 regularizer, which penalizes the discrepancy between the model prediction and the ground truth, constrains the sparsity of model weights, and prevents the model from overfitting.

Compared with the background of the whole QCT scan, the proximal femur only occupies a small number of voxels. To address the data imbalance problem, a Dice loss is employed as the part of the objective function [19]. The Dice similarity coefficient (DSC) measures the amount of agreement between two image regions, as defined in Eq. 1,

$$DSC = \frac{2|g(\hat{y}) \cap y|}{|g(\hat{y})| + |y|} \qquad (1)$$

where $y$ is the ground truth, $g(\hat{y})$ is the binary femur segmentation result obtained from the predicted probability map, and the $|\cdot|$ indicate the number of the voxels. The Dice coefficient between the ground truth and segmented result increases as the deep neural network becomes more powerful, and accuracy of the segmented result increases.

In practice, we use gradient descent-based optimization, which optimizes the weights as the loss becomes smaller and smaller. Hence, we use the Dice loss defined in Eq. 2 as the objective function in pursuit of a higher Dice score:

$$L_{DSC} = 1 - \frac{2|g(\hat{y}) \cap y|}{|g(\hat{y})| + |y|} \tag{2}$$

Our designed V-Net contains 8.6 million parameters in all. Therefore, to accelerate the inference process during the test stage and hasten model convergence in a small number of training epochs, an L2 regularizer is employed in the objective function. The overall objective function is defined in Eq. 3:

$$L_{loss} = L_{DSC} + \alpha \|W\|_2 \tag{3}$$

where the hyper parameter $\alpha$ is an empirical parameter, which is used to balance the loss between Dice loss and L2 loss; W represents all the weights in the 3D V-Net neural network.

The designed V-Net model was implemented in Python using a TensorFlow framework and trained on a workstation with a Tesla Titan V GPU with 12GB GPU memory, an I5 CPU and 16GB RAM. The dropout probability was set as 0.8. The model was trained for 1000 epochs and the weights in the V-Net were optimized using Adam optimizer with a learning rate of 0.0001. Each training epoch cost 129.6, 69.3 and 59.8 seconds for all subjects, and for the male and female cohorts, respectively.

The input to the V-Net was 192×192×32, which indicated that each volume contained 32 cropped QCT slices. However, the number of slices varied among the patients in our dataset. To generate the model prediction of a patient with an arbitrary number of QCT slices, we adopted a sliding window scanning technique which predicted the segmentation from the last axis of the QCT data with an incremental step of 1. That is, if the volume of the QCT image for a specific patient contained 50 QCT slices, the model ran 19 (50-32+1) times to generate the final segmentation result. Typically, to predict the segmentation result for a patient with 60 QCT slices, the proposed model ran 15 seconds on our workstation.

At the training stage, we separately trained our model on the male cohort, the female cohort and all patients. In each cohort, 10% of the samples were used as the test set, and the rest data were used as the training set and validation set. To overcome overfitting and guarantee the generalizability of the model, we fine-tuned our model with a 10-fold cross-validation on the training and validation set. The parameters of the optimized model were acquired according to the highest DSC on the test dataset.

Even though we had images from 397 subjects, the sample size was still insufficient to train a 3D segmentation model. To guarantee the robustness of the model, we randomly rotated the cropped volume by 15 degrees on each axis to augment the data. In addition, during data augmentation, a mirror transform, a brightness transform, a gamma transform, and a Gaussian noise transform were randomly applied to the cropped volume data and the generated samples from each subject were input to our 3D segmentation model for training [20].

### 2.5. Evaluation metrics

We evaluated our segmentation model performance using DSC (Eq. 1), sensitivity (SN) and specificity (SP) metrics. Sensitivity measures the proportion of actual positives that are correctly identified as such. Specificity measures the proportion of actual negatives that are correctly identified as such. As the DSC, sensitivity, and specificity approach 1, the segmentation model results increasingly overlap with the ground truth, and model performance improves.

After generating the segmentation model results, we compared the volume of the proximal femur from the model with the volume of the proximal femur from the ground truth. To measure the volume discrepancy, we examined the $R^2$ score, root mean squared error (RMSE), mean absolute error (MAE), and relative error (RE). A higher $R^2$ score, lower RMSE, lower MAE, and lower RE indicate better model performance.

### 3. Experimental Results

In Figure 4, we juxtaposed the original cropped QCT slices, ground truth and the segmentation model results side-by-side. It can be observed that the contours from our proposed method visually match well with the ground truth.

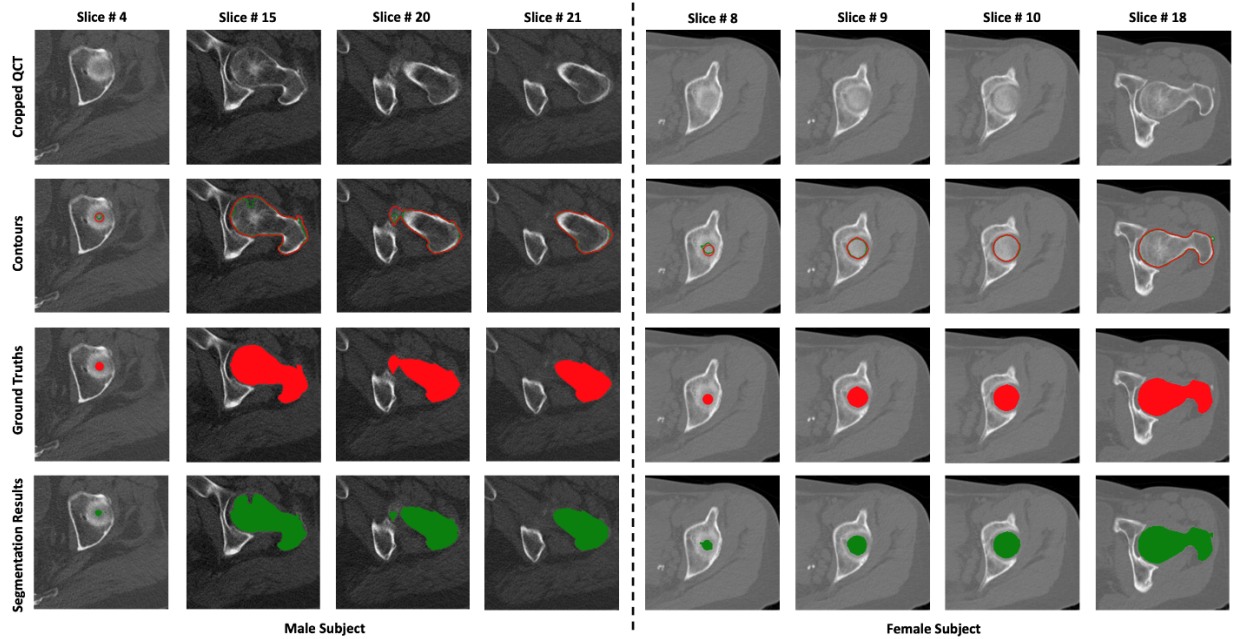

Figure 4. Examples of segmentation model results (green) and the corresponding ground truth (red). In the second row, the green contours represent the segmentation model results, and the red contours indicate the ground truth.

In Figure 5, we visualized the 3D segmentation results of two subjects from the axial, coronal and sagittal axes.

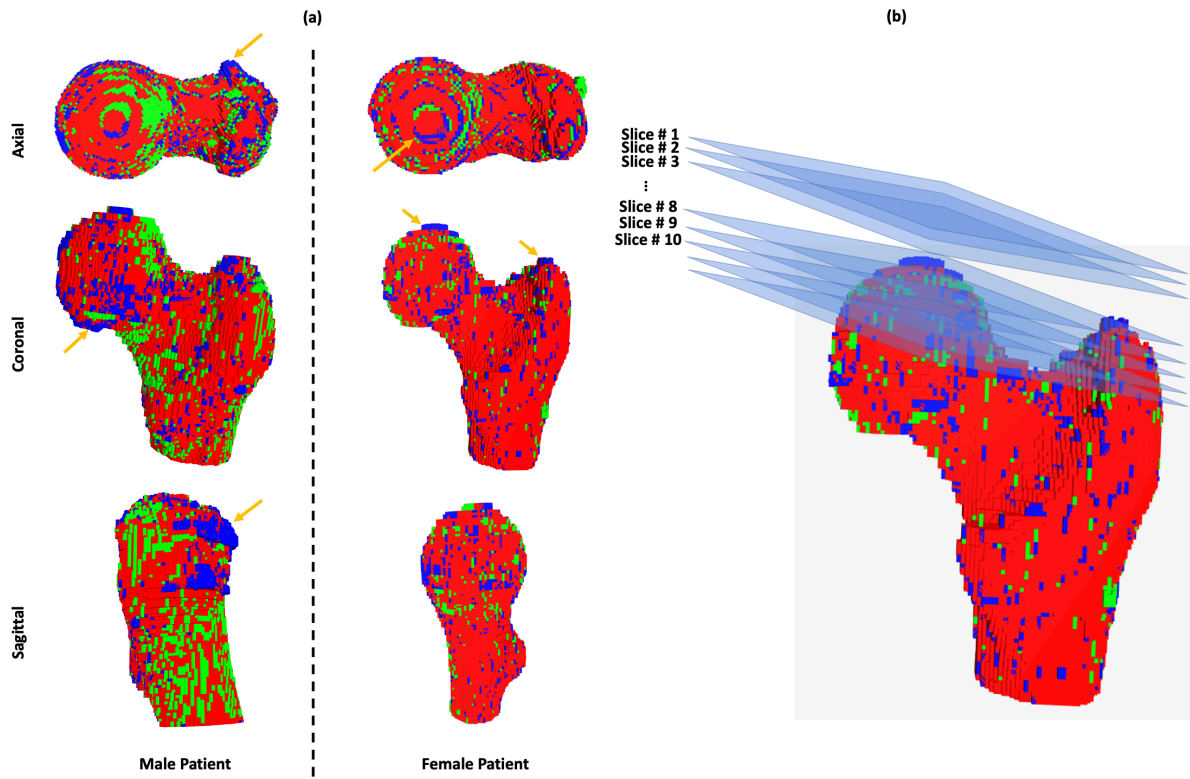

Figure 5. Segmentation results for a female subject and a male subject. The red voxels represent the true positive (TP) voxels, the blue and green voxels indicate the false negative (FN) and false positive (FP) voxels, respectively. The volumes have been resampled with a new spacing of 1mm in all axes.

To quantify the model performance, the mean and standard deviation of DSC, and sensitivity and specificity of the RoI segmentation results for the male cohort, female cohort and all patients were plotted in Figure 6. The quantitative comparison of the femur volumes is shown in Table 1 and Figure 6.

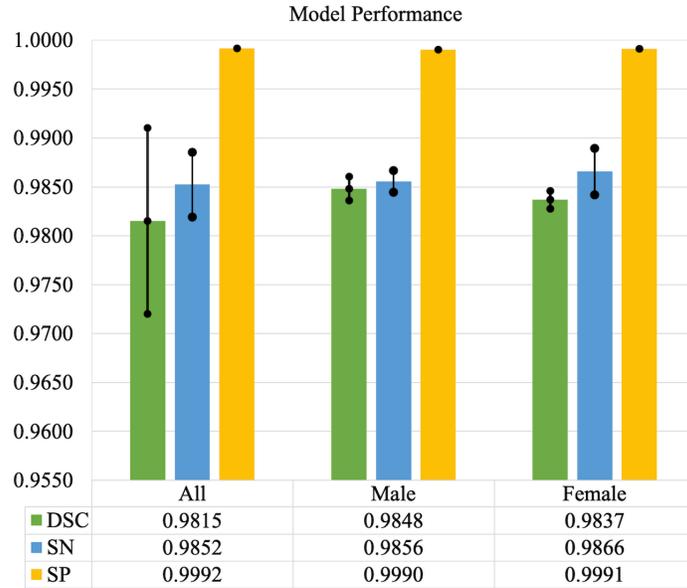

Figure 6. Evaluation metrics for our proposed V-Net model. DSC, Dice similarity coefficient; SN, sensitivity; SP, specificity.

Table 1. Quantitative comparisons between the volumes from the ground truth and our model.

| Metric Cohort | Volume in ground truth (mm$^3$) | MAE (mm$^3$) | RMSE (mm$^3$) | R$^2$ score | RE (%) (min, max) |
|---|---|---|---|---|---|
| All | 89999.59±21325.23 | 901.40 | 1295.66 | 0.9956 | (-2.1476, 4.0612) |
| Male | 105489.94±18343.68 | 658.47 | 883.48 | 0.9977 | (-1.9313, 0.8229) |
| Female | 76832.80±13340.79 | 775.75 | 1361.10 | 0.9896 | (-4.8504, 1.2005) |

*P<0.001 for R$^2$ score in all cohorts (all, male, and female).

To further demonstrate the model performance, we visualized the RE of the volume between the ground truth and our model prediction, as shown in Figure 7.

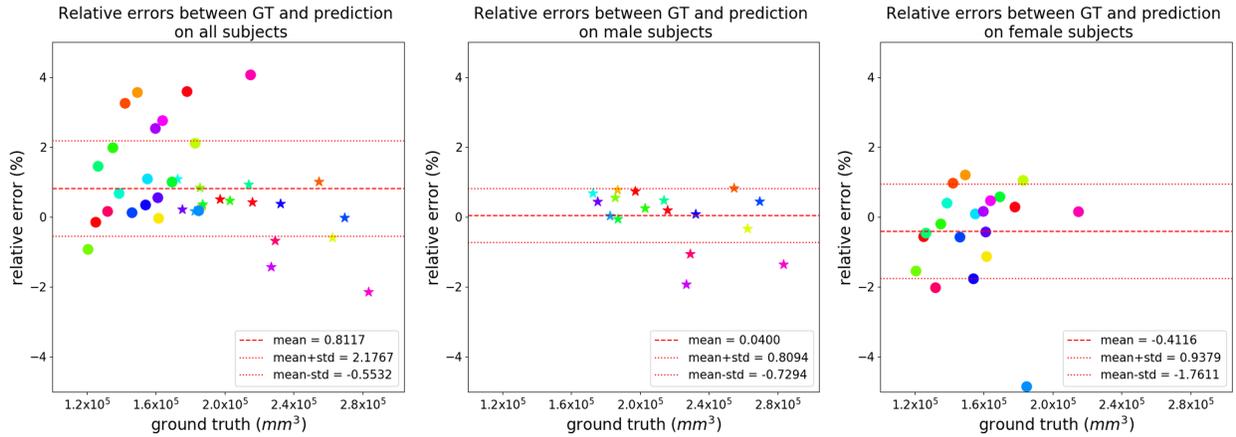

Figure 7. Relative error of femur volume measured by the proposed approach for each patient in each data cohort. The horizontal axis indicates the actual volume of the proximal femur in the QCT images and the vertical axis represents the relative error of the femur volume measured by our prediction compared to the ground truth. The male and female subjects are plotted by star and circle points, receptively. The sample with the same color represents the same subject in each cohort.

## 4. Discussion

### 4.1. Results and Analysis

In our dataset, the slice spacing was much greater than the pixel size so the context information from neighboring slices was not well utilized, reducing the performance of our 3D V-Net. As shown in Figure 5 (b), where the CT slices are indicated by slice 1, slice 2, slice 3 … sequentially, the size of the femoral head within neighboring slices changes dramatically because the slope of the CT scan slices is close to the slope of the femoral head surface. Because of the large slice thickness and the dramatic change in size of the femoral head between slice 1 and slice 2, the image slices could not precisely reflect the shape variation. As a result, most errors occurred where the slope of the bone surface was close to the slope of the CT scan slices and where contrast was low due to low bone density such as in the femoral head. In areas where the slope of the bone surface was far from the slope of the CT scan slices and there was a good deal of contrast, the contours generated by our deep learning model can be used without any changes.

From our perspective, to further improve the performance of our segmentation model, reducing the slice thickness is the most direct approach. However, in most of the slices, the contours generated by our deep learning model can be used without any changes.

According to Table 1, the model performance was much better in the dataset with male patients than the female cohort. This is because in our dataset, the QCT images of female cohort have lower density, which makes segmentation more challenging.

From Table 1 and Figure 7, the discrepancy of the femur volume between the ground truth and prediction in each cohort demonstrated that our model generated more precise results in patients with a smaller femur volume. When the volume was extremely large, the volume from the proposed segmentation model was smaller than the volume from the ground truth.

To further demonstrate the performance of the designed femur segmentation model, we compared the model with other state-of-the-art approaches. The comparison is shown in Table 2.

Table 2. Performance comparison with existing approaches.

| Authors | Method | Data | Sex | Slice Thickness | DSC |
|---|---|---|---|---|---|
| Diogo F. [21] | Active Shape | 148 CT | M&F | 0.625 mm | 0.9400±0.0016 |
| Carballido-Gamio J [6] | Multi-atlas | 210 QCT | M | 2.5mm and 1mm | 0.9760±0.0060 |
| Chen F. [22] | 3D CNN | 150 CT | M&F | 1.32 to 1.85 mm | 0.9688±0.0095 |
| Chang Y. [23] | CRF | 60 CT (120 hemi-hips) | - | 0.45 to 1.2 mm | 0.9490±0.0070 |
| Deniz, C.M. [24] | 3D CNN | 36 3T MR | - | 1.5 mm | 0.9500±0.0200 |
| Present study | 3D CNN | 397 QCT | M&F | 3 mm | **0.9815±0.0009** |

Before comparing our results, we first reviewed existing femur segmentation approaches listed in Table 2. In [21] and [23], the active shape model and conditional random field (CRF) were used to segment the femur from CT images. Both algorithms relied on handcraft features, which severely limited the model performance. In [6], a multi-atlas and registration based methods were employed to extract proximal femur. In [22], a multi-scale fully convolutional neural (FCN) was employed to capture the global features and generate the accurate segmentation results in narrow joint parts. In [24], a U-Net based network and a softmax classifier were used to classify femur voxels into foreground and background. On the contrary, our model generated the probability maps of the femur regions. Compared with the approaches listed in Table 2, we had a large-scale dataset, and the proposed model achieved a higher DSC.

### 4.2. Limitation

Admittedly, there are several limitations about the proposed approach. Model performance was limited by the relatively large slice thickness of the QCT images which reduced reliability of the segmentation results for the femoral head. In addition, the proposed method should be validated by using a larger dataset.

### 5. Conclusion and Future Work

We proposed a V-Net based method to automatically segment the proximal femur in a QCT scan. The proposed model achieved an average Dice score of 0.9815, sensitivity of 0.9852 and specificity of 0.9992. The algorithm we developed can be applied to automatically extract the proximal femur from a QCT image. In addition, it demonstrated a strong potential for clinical use, including the use in finite element analysis part of the proximal femur from QCT images.

The results of this study are promising. With further development to achieve reliable results at the head of the proximal femur, studies of large cohorts of subjects could be analyzed to provide additional insights into evaluating hip fracture risk, the effectiveness of medication and other factors.

In the future, we will improve the performance of the segmentation model to generate precise segmentation results, especially for large shape variations.

**Conflict of interest**

The authors declare no conflicts of interest.

**Acknowledgment**

This research was supported in part by grants from National Institutes of Health, USA (P20GM109036, R01AR069055, U19AG055373, R01AG061917, AR-27065 and M01 RR00585) and NASA Johnson Space Center, USA contracts NNJ12HC91P and NNJ15HP23P. This research was also in part supported by a new faculty startup grant from Michigan Technological University Institute of Computing and Cybersystems.